\documentclass[twocolumn,superscriptaddress,showkeys,english,prl]{revtex4-1}

\usepackage{xcolor} 
\usepackage{ulem} 
\usepackage{changebar}

\usepackage[T1]{fontenc}
\usepackage[latin9]{inputenc}
\usepackage{upgreek,babel,calc,mathrsfs,textgreek,bbm}
\usepackage{amsmath,amssymb,graphicx,xcolor}
\usepackage[unicode=true]{hyperref}

\begin{document}

\title{Optimality in self-organized molecular sorting}

\author{Marco Zamparo}
\thanks{Equal contribution.}
\affiliation{Institute of Condensed Matter Physics and Complex Systems,
Department of Applied Science and Technology, Politecnico di Torino,
Corso Duca degli Abruzzi 24, 10129 Torino, Italy}
\affiliation{Italian Institute for Genomic Medicine c/o Candiolo Cancer Institute, Fondazione del Piemonte per l'Oncologia (FPO), Istituto di Ricovero e Cura a Carattere Scientifico (IRCCS), Candiolo, 10060 Torino, Italy} 
\author{Donatella Valdembri}
\thanks{Equal contribution.}
\affiliation{Department of Oncology, University of Torino School of Medicine,
Candiolo, 10060 Torino, Italy}
\affiliation{Candiolo Cancer Institute, Fondazione del Piemonte per l'Oncologia (FPO),
Istituto di Ricovero e Cura a Carattere Scientifico (IRCCS),
Candiolo, 10060 Torino, Italy}
\author{Guido Serini}
\email{guido.serini@ircc.it}
\affiliation{Department of Oncology, University of Torino School of Medicine,
Candiolo, 10060 Torino, Italy}
\affiliation{Candiolo Cancer Institute, Fondazione del Piemonte per l'Oncologia (FPO),
Istituto di Ricovero e Cura a Carattere Scientifico (IRCCS),
Candiolo, 10060 Torino, Italy}
\author{Igor~V.~Kolokolov}
\email{kolokol@itp.ac.ru}
\affiliation{L.D. Landau Institute for Theoretical Physics,
{142432, Moscow Region, Chernogolovka, Ak. Semenova, 1-A,}
Russia}
\affiliation{National Research University Higher School of Economics,
101000, Myasnitskaya 20, Moscow, Russia}
\author{Vladimir~V.~Lebedev}
\email{lebede@itp.ac.ru}
\affiliation{L.D. Landau Institute for Theoretical Physics,
{142432, Moscow Region, Chernogolovka, Ak. Semenova, 1-A,}
Russia}
\affiliation{National Research University Higher School of Economics,
101000, Myasnitskaya 20, Moscow, Russia}
\author{Luca Dall'Asta}
\email{luca.dallasta@polito.it}
\affiliation{Institute of Condensed Matter Physics and Complex Systems,
Department of Applied Science and Technology,
Politecnico di Torino, Corso Duca degli Abruzzi 24, 10129 Torino, Italy}
\affiliation{Collegio Carlo Alberto, Via Real Collegio 30, 10024 Moncalieri, Italy}
\affiliation{Istituto Nazionale di Fisica Nucleare (INFN), Italy}
\affiliation{Italian Institute for Genomic Medicine c/o Candiolo Cancer Institute, Fondazione del Piemonte per l'Oncologia (FPO), Istituto di Ricovero e Cura a Carattere Scientifico (IRCCS), Candiolo, 10060 Torino, Italy} 
\author{Andrea Gamba}
\email{andrea.gamba@polito.it}
\affiliation{Institute of Condensed Matter Physics and Complex Systems,
Department of Applied Science and Technology, Politecnico di Torino,
Corso Duca degli Abruzzi 24, 10129 Torino, Italy}
\affiliation{Italian Institute for Genomic Medicine c/o Candiolo Cancer Institute, Fondazione del Piemonte per l'Oncologia (FPO), Istituto di Ricovero e Cura a Carattere Scientifico (IRCCS), Candiolo, 10060 Torino, Italy} 
\affiliation{Istituto Nazionale di Fisica Nucleare (INFN), Italy}


\begin{abstract}
We introduce a simple physical picture to explain the process of molecular sorting, 
whereby specific proteins 
are concentrated and
distilled into submicrometric lipid vesicles in eukaryotic cells. 
To this purpose, we formulate a model based on the coupling of spontaneous molecular 
aggregation with vesicle nucleation. Its implications are studied by means of 
a phenomenological theory describing the diffusion of molecules towards multiple 
sorting centers that grow due to molecule absorption and are extracted when 
they reach a sufficiently large size. The predictions of the theory are compared 
with numerical simulations of a lattice-gas 
realization of the model 
and with experimental observations.
The efficiency of the distillation process is found 
to be optimal for intermediate aggregation rates, where the density of sorted 
molecules is minimal and the process obeys simple scaling laws. Quantitative 
measures of endocytic sorting performed in primary endothelial cells are 
compatible with the hypothesis that these optimal conditions are realized 
in living cells.
\end{abstract}

\maketitle

\paragraph{Introduction}
To counter the homogenizing effect of diffusion, eukaryotic cells developed an
elaborate system to sort and distill specific proteins into submicrometric lipid vesicles, that are then
transported towards appropriate intracellular destinations by active mechanisms involving molecular motors~\cite{MN08,SCC+12}.
Molecule sorting takes place on the plasma membrane, on inner membrane bodies (endosomes) and in
the Golgi membrane network. 
Common biochemical principles {involving the action of
specialized proteins that promote membrane bending and fission
{\cite{KCL+14,HBC+10,HK18,MCS+18,BJB+18}}} underlie molecular sorting in these
different locations~{\cite{Tra05,MN08,SCC+12}}. {But can
molecular sorting be understood as a sistemic process, beyond the molecular
detail?} Self-aggregation processes driven by reinforcing feedback loops lead
to the formation of submicrometric domains enriched in specific lipids and
proteins, and are ubiquitous on cell 
membranes~\cite[see Refs.][and references therein]{ZCT+15,HBF18,BBH18,LPR20}. 
Moreover, the formation
of {such} molecular aggregates{, which can be
likened to a phase separation process, that, along with sorted cargo, may
involve several adaptor, membrane-bending and fission-inducing proteins,} has
been observed to precede and induce vesicle nucleation~{\cite{LAD+10,PTC+05}},
and evidences suggest that protein crowding {by itself} can
{drive} membrane bending and vesicle
nucleation~{\cite{SJB08,BHH+15,SSR+12,KR18,Gov18,CAB16}} by making these
processes energetically
favorable~{\cite{Lei86,LA87,BPF11,BBN12,RRT15,RRT15,FS08a}}.
{Altogether,} these observations suggest that sorting may be a
universal process emerging from the coupling of two main components: a) the
self-aggregation of localized protein microdomains, and b) vesicle nucleation.
In this scheme, molecules that diffuse on a membrane can aggregate into localized enriched domains that
grow due to molecule absorption. When a domain reaches a sufficiently large size, its biochemical constituents
locally induce higher membrane curvature and the consequent nucleation and detachment of a small vesicle. The newly generated vesicle
is constitutively enriched in the biochemical factors of the engulfed domain, resulting in a spontaneous distillation process.
Here, 
we formulate a 
phenomenological theory 
of molecular sorting
based on these principles 
and compare its predictions with
numerical simulations 
of a lattice-gas model and quantitative 
measures of the kinetics of endocytic sorting in living cells.
Our analysis suggests 
that higher sorting efficiency is 
realized when 
the number of sorting domains is minimized, a situation taking place for 
intermediate
levels of the
self-aggregation 
strength. 
Our quantitative measures suggest that such 
optimal conditions may be realized in living cells.

\paragraph{Phenomenological Theory}
We 
describe a situation 
where
molecules arrive
on a membrane region, diffuse and aggregate into localized enriched domains, 
and these domains are
removed from the membrane, after reaching a characteristic size~$R_E$, 
through the formation of small separate lipid vesicles.
In this picture,
sorting domains coexist in a statistically stationary state
with a continuously repleted dilute solution, or ``gas'', 
of molecules that diffuse freely on a membrane.
This is reminiscent of two-dimensional diffusion-limited aggregation
(DLA)~\cite{BS95} or the related Hele-Shaw problem~\cite{HS98}.
However, in the advanced stages of DLA large fractal clusters are formed~\cite{BS95},
while in our problem
the presence of the 
cutoff length~$R_E$ 
restricts the domain size, and domain shapes 
remain approximately round.
Here, as in the classical framework of Lifshitz-Slezov (LS) theory~\cite{Sle09,GKL+07}, domains of size $R$ larger
than a critical value $R_{\rm c}$  grow irreversibly by means of the absorption of single molecules
diffusing towards them. This mechanism is expected to provide the dominant contribution to the absorption
dynamics for sufficiently small average molecule density $\bar{n}$ in the gas. For domains with sizes 
much larger than the critical size $R_{\rm c}$, the density $n_0$ near the domain boundary is independent of its size.
When the typical inter-domain distance $L$ is much larger than $R_E$, the difference $\Delta n$ in molecule density between the
regions farther away and closer to the domain boundaries is approximately given by $\bar{n} - n_0 >0$. Contrary
to LS theory, in which domains can grow arbitrarily in time and $\Delta n$ tends to zero
as time grows, here the size of domains removed from the system introduces a cutoff length $R_E \gg R_{\rm c}$, and,
in the statistically stationary regime, $\Delta n$ is kept finite by the continuous influx of particles into the system.

The quasi-static profile of the density 
of freely diffusing molecules
in the vicinity of
a circular domain of size $R$ is
obtained solving the Laplace equation
with Dirichlet boundary conditions:
 \begin{equation}
 n(r) = n_0 + \frac{\ln{r/R}}{\ln{L/R}}\,\Delta n,
 \label{profile}
 \end{equation}
where $r$ is the distance from the domain center. Deviations of the domain shape from 
circularity produce rapidly decaying higher multipole contributions that may be neglected in the main approximation.

The domain grows due to the flux $\Phi_R$ of molecules from the gas, which
can be found by integrating the flux density $-D\,\nabla n$ 
over a circle of radius $r\gg R$: 
\begin{equation}
\label{eq:flux}
 \Phi_R = \left. {2 \pi R D}\, \partial_r n(r) \right|_{r=R}
 = \frac{2 \pi D\Delta n}{\ln({L}/{R})},
 \end{equation}
where $D$ is the diffusion coefficient of isolated molecules.
From~\eqref{eq:flux} one obtains the dynamic equation for domain growth,
$\dot R 
={A_0 D \Delta n}/(R \ln(L/R)),$
where $A_0$ is the area occupied by a molecule in the domain, 
and the domain size $R$ is 
such 
that the domain area
is~$\pi R^2$.

Abstracting from complicated molecular details, the mesoscopic effects of vesicle extraction will be encoded in a single
parameter, the rate $\gamma(R)$ by which domains of size $R$ are removed from the system.
If $N(t,R)\,{\rm d}R$ is the number of domains per unit area with size
between $R$ and $R+{\rm d}R$, the number density $N(t,R)$ satisfies the 
Smoluchowski equation
\begin{equation}
 \frac{\partial N}{\partial t}
 + \frac{\partial}{\partial R}(\dot R\, N  )
 =-\gamma(R)\,N,
 \label{smol}
\end{equation}
A stationary solution of~\eqref{smol} is
\begin{equation}
  N_\mathrm{st} (R)
  =\frac{J R\ln(L/R)}{D\Delta n}
  \exp\left[-\int_0^R\mathrm{d} r\, \frac{r\ln(L/r)\gamma(r)}{A_0D\Delta
  n}\right].
 \label{stat}
\end{equation}
We assume that the extraction rate $\gamma(R)$
is negligible for $R<R_E$ and strongly suppresses $N_\mathrm{st}(R)$ for $R>R_E$, 
where $R_E$ is the characteristic size of the domains
that are extracted from the membrane. 
The factor $J$ in Eq.~(\ref{stat}) is determined 
by noticing that,
in the stationary regime, the average flux 
$\int \Phi_R N_{\rm st}(R)\,\mathrm{d} R$
must equate 
the incoming flux of molecules per unit area $\phi$ 
(one of the control parameters of the theory), thus
giving $J \sim
\phi/R_E^2$.
In the region $R<R_E$ where $\gamma(R)$ is negligible,
Eq.~\eqref{stat} shows that the distribution $N_\mathrm{st}(R)$ has a universal
behavior characterized by a linear growth with logarithmic corrections. 
The present phenomenological approach is applicable if 
the inequality $R_E^2\gg A_0$ is satisfied.
This condition also justifies the quasi-static approach leading to Eq.~(\ref{profile}). 

The efficiency of the sorting process can be measured in terms of 
the average residence time $\bar{T}$ of a molecule on the 
membrane system. 
For absorbing domains, this 
is the sum of the average time~$\bar{T}_{\rm f}$
required by the molecule to reach a domain by free diffusion and be absorbed, 
and the average time $\bar{T}_{\rm d}$ spent 
inside that domain 
until the extraction event. For evenly distributed domains, the first contribution~$\bar{T}_{\rm f}$
is inversely proportional to the average number $N_\mathrm{d}$ of 
domains per unit area, where
$N_{\rm d} = \int \mathrm{d} R\, N_{\rm st}(R)\sim {\phi}/({D \Delta n}),$
giving
$\bar{T}_{\rm f}
\sim  {1}/({D N_{\rm d}}) \sim {\Delta n}/{\phi}.$
In its turn, the average time spent by a molecule in a domain can be estimated as
$\bar T_{\rm d}\sim {R_E^2 }/({A_0 \Phi_R})
\sim {R_E^2}/({D A_0 \Delta n}),$
where~\eqref{eq:flux} was used.

The rate of formation of new domains can be estimated as
${\mathrm{d}
N_\mathrm{d}}/{\mathrm{d}t}
= C D \,\bar n^2,$
where $C$ is a dimensionless quantity characterizing the efficiency of absorption of 
single molecules by the germ of
a domain.  In the stationary condition this rate
is equal to $N_\mathrm{d} /\bar T_\mathrm{d}$, therefore
 \begin{equation}
 \bar n\sim \left(\frac{\phi A_0}{CD
 R_E^2}\right)^{1/2}.
 \label{density}
 \end{equation}
Assuming $n_0\lesssim \Delta n$ we get $\Delta n\sim \bar n$ and then
 \begin{equation}
 \bar T_\mathrm{d} \sim
 C^{1/2}\frac{R_E^3}{(D\phi)^{1/2}A_0^{3/2}},
 \quad \bar T_\mathrm{f}\sim
 C^{-1/2} \frac{A_0^{1/2}}{(D\phi)^{1/2}R_E}.
 \label{barts}
 \end{equation}
The sum $\bar T=\bar T_{\mathrm d}+ \bar T_f$, as a function of $C$, has a minimum in $C\sim A_0^2/R_E^4\ll1$, where
\begin{eqnarray}
\bar T_{\rm f}\sim\bar T_{\rm d}\sim
\frac{R_E}{(DA_0)^{1/2}\phi^{1/2}},
\label{optimum} \\
 \bar n\sim \Delta n \sim
\frac{\phi^{1/2} R_E}{(DA_0)^{1/2}}.
 \label{deltann}
\end{eqnarray}
Therefore, the scaling relations (\ref{optimum},\ref{deltann}) identify the dynamical regime
in which molecular sorting is most efficient. 
The density of molecules accumulated in the domains is
\begin{equation}
\rho_{\mathrm d} \sim N_{\mathrm d}
R_E^2/A_0\sim C^{1/2}\frac{\phi^{1/2}R_E^3}{D^{1/2}A_0^{3/2}}.
\label{dommol}
\end{equation}
\begin{figure*}[ht]
      \begin{center}
      ~\hspace{-0.2cm}{\includegraphics[width=2.08\columnwidth]{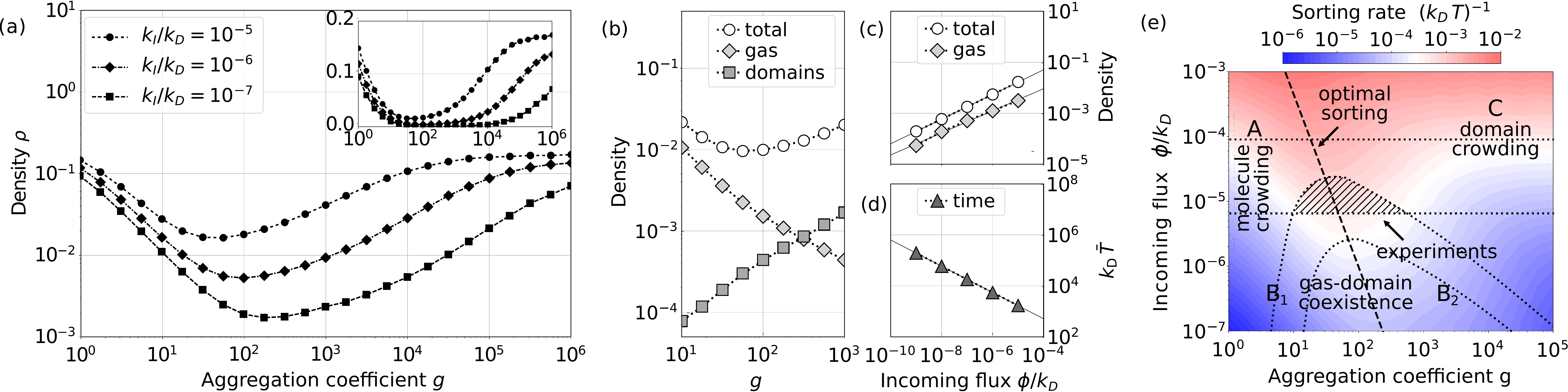}} 
      \end{center}
  \caption{(a) Molecule density $\rho$
  (time average in the statistically stationary state) as a function of the aggregation
  coefficient~$g$, for $k_I / k_D=10^{-5},\,10^{-6},\,10^{-7}$.
  Inset: same quantities in log-linear scale.
(b)~Partial molecule densities as functions of the aggregation
  coefficient $g$. The density of the gas component decreases, 
  while the number of molecules in the interior of
  sorting domains increases 
  for 
  growing
  $g$ at fixed $\phi / k_D = 10^{-8}$;
  as a consequence, the total molecule density has a minimum at an 
  intermediate value of $g$.
  (c,d)~Numerical evidence of scaling relations.
   Straight lines are fitted with the laws
  $\rho \sim\phi^{a}$, $\bar{n}\sim\phi^{b}$, 
  and $\bar{T}\sim\phi^{-{c}}$ with $a=0.48\pm0.01$, $b=0.46\pm0.02$, 
  and $c=0.52\pm0.01$.
(e)~Nondimensionalized sorting rate $(k_D\bar T)^{-1}$ as a function of the aggregation coefficient $g$
 and of the 
 nondimensionalized incoming flux~$\phi/k_D$. 
  At fixed~$\phi$, faster distillation takes place for intermediate 
  values of~$g$, thus showing the existence of a region of parameter space
  where sorting is optimal. In this qualitative phase diagram, A,~C~are 
  high-density phases, characterized by
  molecule crowding and domain crowding, respectively; B~is a low-density phase
  where 
  the average domain size is much less than the average interdomain separation and
  sorting domains coexist with freely-diffusing molecules. 
  The dashed line marks maximal sorting at
  fixed flux $\phi$, and divides the B region into regions of less~(B$_1$) 
  and more dilute gas~(B$_2$).
 Model parameters compatible with experimental
     values of molecule density and flux
     are shown as a shaded area. }
  \label{fig:flux}
\end{figure*}
Thus, also the total density $\rho=\bar n+\rho_{\mathrm d}$ has a minimum for $C\sim A_0^2/R_E^4$,
and the minimal value of $\rho$ is again determined by the estimate~(\ref{deltann}).

\paragraph{Lattice-gas model and numerical results}
To further explore the role of molecule self-aggregation in the distillation process, and to probe the behavior
of the sorting process over a wide range of parameter values, we introduce here a minimal lattice-gas model of molecular
sorting,
without any attempt at a
complete description of the complex biochemical and physical details implied in the process of vesicle budding and removal.

We represent the lipid membrane as a two-dimensional square lattice 
with periodic boundary conditions, where 
each lattice site 
can
host at most a single molecule~\cite{BSG+07}. 
The system evolves
according to a Markov process that
comprises the following three elementary mechanisms: {1)}~Molecules from an infinite reservoir arrive and are inserted on
empty sites with rate~$k_I$. {2)} Then, molecules can perform diffusive jumps to an empty neighboring site with rate
$k_D/g^{\#\mathrm{nn}}$, where $g>1$ is a dimensionless aggregation coefficient and $\#\mathrm{nn}$ is the number of 
molecules 
neighboring the site originally occupied by
the jumping molecule. {3)}  Finally, molecules are extracted from the system by the
simultaneous removal 
of all connected molecule clusters, if any, that contain a completely filled square
of linear size $\ell$, with $\ell^2 \sim R_E^2/A_0$ 
(for a formal mathematical definition see SM).
The stationary properties of the model depend on only two parameters, the ratio
$k_I/k_D$ and the aggregation coefficient $g$. 
In what follows, areas are measured in units of a
lattice site, therefore $A_0=1$, and the particle densities $\rho$, $n$ 
are dimensionless quantities.

The statistically stationary state of the model was investigated numerically.
Fig.~\ref{fig:flux}(a) shows that the 
stationary density of molecules~$\rho$
is 
low for intermediate values of~$g$, where
a dilute gas of free molecules coexists with growing 
domains. 
In this region, the fraction of free molecules decreases as $g$ is increased, 
and 
the total
molecule density has 
a minimum. 
The neighborhood of
this minimum corresponds to the region previously found 
from the analysis of the
phenomenological theory, which is likely 
the most interesting from the biological point of view.
In Fig.~\ref{fig:flux}(b) the total density $\rho$ is decomposed into the contributions of freely diffusing molecules and of the molecules which
are part of sorting domains. When $g$~is increased, the density of freely diffusing
molecules decreases, while 
the number of molecules in the domains increases,
leading to
the appearance of a minimum of~$\rho$ (Fig.~\ref{fig:flux}(b), white circles)
at intermediate values of the aggregation coefficient~$g$. In this region, 
we computed numerical scaling relations with
respect to the incoming flux per unit site $\phi=k_I(1-\rho)$, finding good agreement with the theoretical 
predictions~(\ref{optimum}--\ref{dommol})
(Fig.~\ref{fig:flux}\hbox{(c)--(d)} and figure legend).

To characterize the efficiency of the sorting process we computed numerically the 
sorting rate
$\bar T^{-1}=\phi/\rho$ 
(see SM and Ref.~\cite{ZAG19}) in terms of the physically
meaningful parameters $\phi$~and~$g$~\footnote{Since $\phi$ is a
monotonically increasing function of the insertion 
rate~$k_I$~(see SM), by
a change of variable~$\phi$ (which is directly observable) can be used as a
control parameter in the place of~$k_I$ (which is not observable).}.
Fig.~\ref{fig:flux}(e) shows that ${\bar T}^{-1}$ increases monotonically with~$\phi$, 
and that it exhibits a maximum as a function of $g$ at fixed
$\phi$ (the dashed line in~Fig.~\ref{fig:flux}(e) marks the position of these maxima).  
In~the
optimal sorting region located around the maxima of ${\bar T}^{-1}$
distillation of molecular factors is most efficient. 

The numerical evidence of a region of optimal sorting is 
in agreement with the phenomenological theory.
This can be 
seen by considering that, in the framework of the numerical scheme, the efficiency $C$
of absorption of single molecules increases monotonically with $g$. Then the existence of the
maximum of the sorting rate 
$\bar T^{-1}$ 
and of the minimum of the density $\rho$ observed in the
numerical modeling appears as a natural consequence of the phenomenological theory. 
The
contrasting behavior of the density of particles in the gas and 
in the domains (Fig.~\ref{fig:flux}(b)) agrees
with~Eqs.~(\ref{density},\ref{dommol}).

Along with the residence time $\bar T$, which is a property of the stationary state, 
we considered also 
the characteristic adaptation time $T_\mathrm{ad}$ needed by the membrane system 
to 
approach the stationary state after the 
sudden onset of a nonzero external stimulus. 
Numerical simulations show that $T_\mathrm{ad}$  
is directly correlated to $\bar T$ (see SM). 
Therefore,
parameter values 
that correspond to
optimal sorting in the stationary state 
are also those that
provide faster 
response to changing environmental signals.

\begin{figure}[t]
  \begin{center}
      \vspace{0.0cm}\resizebox{\columnwidth}{!}{\includegraphics{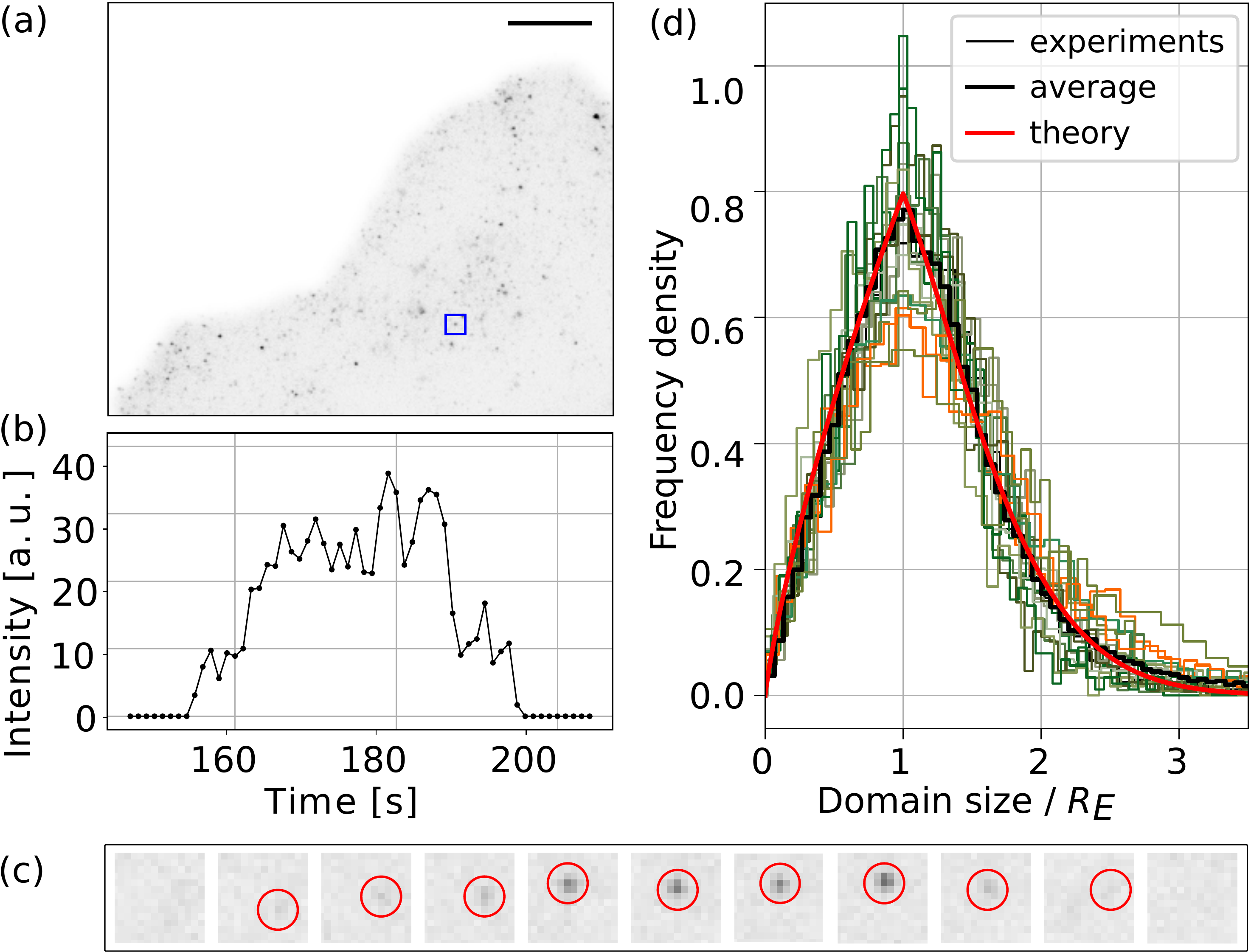}}
  \end{center}
  \caption{Sorting domains on the cell plasmamembrane.
  (a)~TIRF micrograph of LDL sorting domains;
  scale bar:
  10\,\textmu m;
  blue square: sample region of interest.
  (b) Time course of fluorescence intensity in the region of interest. 
  (c) Snapshots of the region of
  interest separated by $\Delta t =5\,\mathrm{s}$ intervals, starting from
  $t=150\,\mathrm{s}$; 
  red circles are centered around an automatically identified growing domain.
  (d) Frequency density of domain sizes: comparison of 
  a fit of the theoretical curve (\ref{stat}) to
  data 
  from 
  18 cells, pooled from 
  two independent
  experiments.
  The domain size distribution computed
   from the complete experimental dataset is 
   represented by the black line.
Domain sizes were estimated as described in the text and SM.  }
  \label{fig:endocytosis}
\end{figure}

\paragraph{Comparison with experimental data}
\label{sec:exp}
It is interesting to check how experimental data 
compare with our general physical theory.
Low-density lipoproteins (LDL) bind  
to 
LDL receptors (LDLR) on
the plasma membrane
in a 1:1 ratio, diffuse laterally,
aggregate and are 
internalized in 
endocytic
vesicles (see Refs.~\onlinecite{GWB81}, \onlinecite{MN08} and SM, 
which includes additional 
Refs.~\onlinecite{Gil76,BG86,HPH+00,Sch16,IPM86,EBV+04,OSW+19,WW14,KCT+18,Mod06,Lin98,DKL+07,DVC+11,ASB+15,KBJ83,WS81,Mea83}).
They 
provide a convenient experimental system whose behavior can be compared 
with the theory. 
We 
performed experiments of endocytic sorting of LDL 
on 
primary human endothelial cells
incubated with LDL particles for 1~hour prior to imaging to allow them
to reach a stationary state
(see~\cite{MCP+16} and SM for details). 
LDL particles tagged with 
Alexa Fluor 488 green fluorescent dye  
were employed (SM).
The local density of fluorescently-tagged molecules was quantified using 
total internal reflection fluorescence (TIRF) microscopy
(Fig.~\ref{fig:endocytosis}(a)), 
which allows to constrain the analysis
to a thin layer of approximately $100\,\mathrm{nm}$ from the plasma membrane~\cite{PK18}. 
In these
experimental conditions it was possible to observe a large number of 
endocytic events
including the 
formation and detachment
of LDL-enriched vesicles 
(Fig.~\ref{fig:endocytosis}(b,c)).
Growing fluorescently-tagged domains
were identified by automated image analysis.
Assuming that LDL particles
are distributed with approximately constant, uniform probability 
on the surface of sorting domains, and neglecting curvature corrections,
the cumulative fluorescence intensity collected
from a given domain is expected to be 
approximately proportional, in average, to the area
of the domain.
In order 
to fix the conversion factor,
we assumed that the typical
fluorescence intensity 
reached by growing domains just before their 
extraction corresponds to the 
size
of mature 
endocytic
vesicles,
$R_E \sim 100\,\mathrm{nm}$~\cite{KR18}~(see~SM). 
Histograms of domain sizes 
(Fig.~\ref{fig:endocytosis}(d)) 
show an 
approximately linear dependence of the frequency density on domain size for 
radii $R<R_E$,
compatibly with the behavior predicted by Eq.~(\ref{stat}). 
Assuming 
$\gamma(R)=0$ for  
${R<R_E}$ and $\gamma(R)=\gamma_0$ for $R>R_E$,
and fitting~$\gamma_0$, Eq.~(\ref{stat})
can be adapted to the
experimental data also for 
$R>R_E$
(Fig.~\ref{fig:endocytosis}(d)). 

We measured LDL\, surface\, density 
${\bar\rho = 1.7\cdot 10^{-2}}$
($\sigma_\rho=0.8\cdot 10^{-2}$)
and flux
${\bar\phi = 2.8 \cdot 10^{-5}}$ 
($\sigma_\phi=1.8\cdot 10^{-5}$)
expressed, respectively, as fraction of the cell surface covered by sorting domains, 
and fraction of the cell surface extracted per second
(see SM), 
compatibly with previous observations~\cite{GWB81}. 
The microscopic rate $k_D$ can be 
estimated as
$D/A_0$, where 
$D=3\cdot 10^{-3}\,\upmu\mathrm{m}^2/\mathrm{s}$ 
($\sigma_D=1.5\cdot 10^{-3}\,\upmu\mathrm{m}^2/\mathrm{s}$) 
is the 
lateral diffusivity of LDL molecules~\cite{BW82,GWB81}, 
and 
$A_0 \sim 3\cdot 10^{-3} \,\upmu\mathrm{m}^2$~\footnote{Single 
molecule measurements 
show that endocytic vesicles host 
$\pi R_E^2/A_0\sim 10$ LDL molecules~\cite{ABG77}.}.
The relation $\bar\phi = k_I (1-\bar\rho)\simeq k_I$ allows then to estimate 
$k_I/k_D\sim 10^{-5}$.
Fig.~\ref{fig:flux}(a) shows that for such parameter values, experimentally measured densities are attained in the physical model
in the vicinity of the minima of the density $\rho$, i.e in the optimal region.
On the phase diagram Fig.~\ref{fig:flux}(e)
the 
$g$,~$\phi$ pairs 
compatible with the experiments
are
found 
at the intersection of the 
regions comprised between 
the 
numerically computed 
curves of equation  
$\rho=\bar\rho \pm \sigma_\rho$ and $\phi/k_D=(\bar\phi\pm\sigma_\phi)
     \,A_0\,(D\mp\sigma_D)^{-1}$ (dotted lines in Fig.~\ref{fig:flux}(e)), 
and 
are
situated in a neighborhood of the optimal region. 

\paragraph{Discussion}
The observation of 
eukaryotic cells 
by fluorescence microscopy
shows
the hectic and apparently
chaotic traffic of a myriad of submicrometric vesicles that transport lipids 
and proteins to 
disparate subcellular locations~\cite{SDN+00}.
This restless movement takes place at significant energy cost, suggesting 
that it must be deeply relevant for cell life.
Actually, 
to perform its 
vital tasks, such as feeding on nutrients,
proliferating, migrating, and forming complex 
multicellular tissues, 
the cell 
has first of all to break its
original
symmetry~\cite{WL03}:
in the process,
each 
region of 
its outer 
and inner 
membranes
becomes 
endowed with a 
specific chemical identity,
that allows 
it to perform 
its peculiar functions~\cite{RM14a}.
Vesicular traffic 
creates and sustains 
this broken-symmetry state: vesicles enriched in
specific molecular factors are 
continuosly
delivered to appropriate membrane regions 
to maintain their 
biochemical identity and to contrast
the homogenizing effect of diffusion~\cite{SEM11}. 
But 
what are the physical bases of this universal distillation mechanism? 
We~have proposed here 
a scenario whereby 
molecular sorting
emerges 
from the coupling of two 
spontaneous 
processes:
a)~phase separation induced by
molecular self-aggregation~\cite{GCT+05,GKL+07,ZCT+15,HBF18,BBH18,LPR20}, 
and~b)~vesicle nucleation.
This view has solid bases in the physical chemistry of vesicular traffic, as
several mechanisms have been identified that link the formation
of molecular aggregates to the 
induction of membrane bending and 
vesicle nucleation~\cite{Lei86,LA87,HBC+10,BPF11,BBN12,SSR+12,KCL+14,BHH+15,KR18,HK18,MCS+18,BJB+18}.
From this general scenario, 
the following picture emerges: a
continuously repleted gas of molecules diffusing towards multiple sorting centers,
that grow due to molecular absorption and 
lead
to the formation of  vesicles in which a
higher-than-average concentration of the given molecular factor has been
distilled.  
The emergence of an optimal sorting regime at intermediate values of the 
aggregation strength then follows as a nontrivial effect
of the physics of diffusion-limited aggregation 
on cell membranes, 
pointing
out at the central role of
self-aggregation in ordering the system. 
In our experiments 
coalescence of domains is rarely observed (see SM), suggesting that 
domains grow mainly by 
the absorption of
laterally diffusing molecules from the surrounding molecule gas. 
The opposite
regime, where domains mainly grow by coalescence, 
has been proposed to describe the formation of exit sites in 
the endoplasmic reticulum~\cite{HWK+08},
suggesting that 
it would be interesting to develop a more general framework
where both molecular aggregation and domain coalescence are relevant.
Although here for simplicity we have considered the 
distillation of a single molecular factor, 
it is 
immediately evident that distinct 
clans 
of molecules
endowed with high intra-clan affinity can separately aggregate 
in distinct enriched domains
and be sorted in parallel.
Clans of molecules 
that participate
in common networks of 
reinforcing 
catalytic 
feedback loops 
are obvious candidates for the spontaneous formation of such 
enriched domains~\cite{GCT+05,GKL+07,ZCT+15,HBF18,BBH18,LPR20}.
By~measuring the density of sorting domains and the sorting flux, we observed
that 
LDL endocytosis in
primary human endothelial cells kept in 
steady-state 
conditions 
takes place
close to the optimal regime.
It~is then tempting to speculate that an evolutionary constraint may have led the
proteins responsible for the distillation process to tune their activity around
optimality, 
as maximal sorting efficiency 
may have provided
selective advantage in terms
of faster adaptation to rapidly varying environmental conditions.

\begin{acknowledgments}MZ, LDA and AG thank Carlo Cosimo Campa, Jean Piero Margaria,
Emiliano Descrovi
and Emilio Hirsch for useful discussions.
Numerical calculations have been made possible through a CINECA-INFN agreement 
providing access to resources on MARCONI at CINECA.
The experimental research described in this work was 
supported by funding from Fondazione AIRC IG grants \#21315 (to GS) and \#20366 (to DV).
\end{acknowledgments}


\begin{thebibliography}{62}%
\makeatletter
\providecommand \@ifxundefined [1]{%
 \@ifx{#1\undefined}
}%
\providecommand \@ifnum [1]{%
 \ifnum #1\expandafter \@firstoftwo
 \else \expandafter \@secondoftwo
 \fi
}%
\providecommand \@ifx [1]{%
 \ifx #1\expandafter \@firstoftwo
 \else \expandafter \@secondoftwo
 \fi
}%
\providecommand \natexlab [1]{#1}%
\providecommand \enquote  [1]{``#1''}%
\providecommand \bibnamefont  [1]{#1}%
\providecommand \bibfnamefont [1]{#1}%
\providecommand \citenamefont [1]{#1}%
\providecommand \href@noop [0]{\@secondoftwo}%
\providecommand \href [0]{\begingroup \@sanitize@url \@href}%
\providecommand \@href[1]{\@@startlink{#1}\@@href}%
\providecommand \@@href[1]{\endgroup#1\@@endlink}%
\providecommand \@sanitize@url [0]{\catcode `\\12\catcode `\$12\catcode
  `\&12\catcode `\#12\catcode `\^12\catcode `\_12\catcode `\%12\relax}%
\providecommand \@@startlink[1]{}%
\providecommand \@@endlink[0]{}%
\providecommand \url  [0]{\begingroup\@sanitize@url \@url }%
\providecommand \@url [1]{\endgroup\@href {#1}{\urlprefix }}%
\providecommand \urlprefix  [0]{URL }%
\providecommand \Eprint [0]{\href }%
\providecommand \doibase [0]{http://dx.doi.org/}%
\providecommand \selectlanguage [0]{\@gobble}%
\providecommand \bibinfo  [0]{\@secondoftwo}%
\providecommand \bibfield  [0]{\@secondoftwo}%
\providecommand \translation [1]{[#1]}%
\providecommand \BibitemOpen [0]{}%
\providecommand \bibitemStop [0]{}%
\providecommand \bibitemNoStop [0]{.\EOS\space}%
\providecommand \EOS [0]{\spacefactor3000\relax}%
\providecommand \BibitemShut  [1]{\csname bibitem#1\endcsname}%
\let\auto@bib@innerbib\@empty
\bibitem [{\citenamefont {Mellman}\ and\ \citenamefont {Nelson}(2008)}]{MN08}%
  \BibitemOpen
  \bibfield  {author} {\bibinfo {author} {\bibfnamefont {I.}~\bibnamefont
  {Mellman}}\ and\ \bibinfo {author} {\bibfnamefont {W.~J.}\ \bibnamefont
  {Nelson}},\ }\href {https://doi.org/10.1038/nrm2525} {\bibfield  {journal}
  {\bibinfo  {journal} {Nat. Rev. Mol. Cell Biol.}\ }\textbf {\bibinfo {volume}
  {9}},\ \bibinfo {pages} {833} (\bibinfo {year} {2008})}\BibitemShut {NoStop}%
\bibitem [{\citenamefont {Sigismund}\ \emph {et~al.}(2012)\citenamefont
  {Sigismund}, \citenamefont {Confalonieri}, \citenamefont {Ciliberto},
  \citenamefont {Polo}, \citenamefont {Scita},\ and\ \citenamefont {{Di
  Fiore}}}]{SCC+12}%
  \BibitemOpen
  \bibfield  {author} {\bibinfo {author} {\bibfnamefont {S.}~\bibnamefont
  {Sigismund}}, \bibinfo {author} {\bibfnamefont {S.}~\bibnamefont
  {Confalonieri}}, \bibinfo {author} {\bibfnamefont {A.}~\bibnamefont
  {Ciliberto}}, \bibinfo {author} {\bibfnamefont {S.}~\bibnamefont {Polo}},
  \bibinfo {author} {\bibfnamefont {G.}~\bibnamefont {Scita}}, \ and\ \bibinfo
  {author} {\bibfnamefont {P.~P.}\ \bibnamefont {{Di Fiore}}},\ }\href
  {\doibase 10.1152/physrev.00005.2011} {\bibfield  {journal} {\bibinfo
  {journal} {Physiol Rev}\ }\textbf {\bibinfo {volume} {92}},\ \bibinfo {pages}
  {273} (\bibinfo {year} {2012})}\BibitemShut {NoStop}%
\bibitem [{\citenamefont {Kozlov}\ \emph {et~al.}(2014)\citenamefont {Kozlov},
  \citenamefont {Campelo}, \citenamefont {Liska}, \citenamefont {Chernomordik},
  \citenamefont {Marrink},\ and\ \citenamefont {McMahon}}]{KCL+14}%
  \BibitemOpen
  \bibfield  {author} {\bibinfo {author} {\bibfnamefont {M.~M.}\ \bibnamefont
  {Kozlov}}, \bibinfo {author} {\bibfnamefont {F.}~\bibnamefont {Campelo}},
  \bibinfo {author} {\bibfnamefont {N.}~\bibnamefont {Liska}}, \bibinfo
  {author} {\bibfnamefont {L.~V.}\ \bibnamefont {Chernomordik}}, \bibinfo
  {author} {\bibfnamefont {S.~J.}\ \bibnamefont {Marrink}}, \ and\ \bibinfo
  {author} {\bibfnamefont {H.~T.}\ \bibnamefont {McMahon}},\ }\href
  {https://doi.org/10.1016/j.ceb.2014.03.006} {\bibfield  {journal} {\bibinfo
  {journal} {Current Opinion in Cell Biology}\ }\textbf {\bibinfo {volume}
  {29}},\ \bibinfo {pages} {53} (\bibinfo {year} {2014})}\BibitemShut {NoStop}%
\bibitem [{\citenamefont {Hurley}\ \emph {et~al.}(2010)\citenamefont {Hurley},
  \citenamefont {Boura}, \citenamefont {Carlson},\ and\ \citenamefont
  {Ró{\.z}ycki}}]{HBC+10}%
  \BibitemOpen
  \bibfield  {author} {\bibinfo {author} {\bibfnamefont {J.~H.}\ \bibnamefont
  {Hurley}}, \bibinfo {author} {\bibfnamefont {E.}~\bibnamefont {Boura}},
  \bibinfo {author} {\bibfnamefont {L.-A.}\ \bibnamefont {Carlson}}, \ and\
  \bibinfo {author} {\bibfnamefont {B.}~\bibnamefont {R{\'o}{\.z}ycki}},\ }\href
  {\doibase 10.1016/j.cell.2010.11.030} {\bibfield  {journal} {\bibinfo
  {journal} {{Cell}}\ }\textbf {\bibinfo {volume} {143}},\ \bibinfo {pages}
  {875} (\bibinfo {year} {2010})}\BibitemShut {NoStop}%
\bibitem [{\citenamefont {Haucke}\ and\ \citenamefont {Kozlov}(2018)}]{HK18}%
  \BibitemOpen
  \bibfield  {author} {\bibinfo {author} {\bibfnamefont {V.}~\bibnamefont
  {Haucke}}\ and\ \bibinfo {author} {\bibfnamefont {M.~M.}\ \bibnamefont
  {Kozlov}},\ }\href {https://doi.org/10.1242/jcs.216812} {\bibfield  {journal}
  {\bibinfo  {journal} {{J Cell Sci}}\ }\textbf {\bibinfo {volume} {131}}
  (\bibinfo {year} {2018})}\BibitemShut {NoStop}%
\bibitem [{\citenamefont {Mettlen}\ \emph {et~al.}(2018)\citenamefont
  {Mettlen}, \citenamefont {Chen}, \citenamefont {Srinivasan}, \citenamefont
  {Danuser},\ and\ \citenamefont {Schmid}}]{MCS+18}%
  \BibitemOpen
  \bibfield  {author} {\bibinfo {author} {\bibfnamefont {M.}~\bibnamefont
  {Mettlen}}, \bibinfo {author} {\bibfnamefont {P.-H.}\ \bibnamefont {Chen}},
  \bibinfo {author} {\bibfnamefont {S.}~\bibnamefont {Srinivasan}}, \bibinfo
  {author} {\bibfnamefont {G.}~\bibnamefont {Danuser}}, \ and\ \bibinfo
  {author} {\bibfnamefont {S.~L.}\ \bibnamefont {Schmid}},\ }\href
  {http:/dx.doi.org/10.1146/annurev-biochem-062917-012644} {\bibfield
  {journal} {\bibinfo  {journal} {Annual review of biochemistry}\ }\textbf
  {\bibinfo {volume} {87}},\ \bibinfo {pages} {871} (\bibinfo {year}
  {2018})}\BibitemShut {NoStop}%
\bibitem [{\citenamefont {Bassereau}\ \emph {et~al.}(2018)\citenamefont
  {Bassereau}, \citenamefont {Jin}, \citenamefont {Baumgart}, \citenamefont
  {Deserno}, \citenamefont {Dimova}, \citenamefont {Frolov}, \citenamefont
  {Bashkirov}, \citenamefont {Grubmüller}, \citenamefont {Jahn}, \citenamefont
  {Risselada}, \citenamefont {Johannes}, \citenamefont {Kozlov}, \citenamefont
  {Lipowsky}, \citenamefont {Pucadyil}, \citenamefont {Zeno}, \citenamefont
  {Stachowiak}, \citenamefont {Stamou}, \citenamefont {Breuer}, \citenamefont
  {Lauritsen}, \citenamefont {Simon}, \citenamefont {Sykes}, \citenamefont
  {Voth},\ and\ \citenamefont {Weikl}}]{BJB+18}%
  \BibitemOpen
  \bibfield  {author} {\bibinfo {author} {\bibfnamefont {P.}~\bibnamefont
  {Bassereau}}, \bibinfo {author} {\bibfnamefont {R.}~\bibnamefont {Jin}},
  \bibinfo {author} {\bibfnamefont {T.}~\bibnamefont {Baumgart}}, \bibinfo
  {author} {\bibfnamefont {M.}~\bibnamefont {Deserno}}, \bibinfo {author}
  {\bibfnamefont {R.}~\bibnamefont {Dimova}}, \bibinfo {author} {\bibfnamefont
  {V.~A.}\ \bibnamefont {Frolov}}, \bibinfo {author} {\bibfnamefont {P.~V.}\
  \bibnamefont {Bashkirov}}, \bibinfo {author} {\bibfnamefont {H.}~\bibnamefont
  {Grubmüller}}, \bibinfo {author} {\bibfnamefont {R.}~\bibnamefont {Jahn}},
  \bibinfo {author} {\bibfnamefont {H.~J.}\ \bibnamefont {Risselada}}, \bibinfo
  {author} {\bibfnamefont {L.}~\bibnamefont {Johannes}}, \bibinfo {author}
  {\bibfnamefont {M.~M.}\ \bibnamefont {Kozlov}}, \bibinfo {author}
  {\bibfnamefont {R.}~\bibnamefont {Lipowsky}}, \bibinfo {author}
  {\bibfnamefont {T.~J.}\ \bibnamefont {Pucadyil}}, \bibinfo {author}
  {\bibfnamefont {W.~F.}\ \bibnamefont {Zeno}}, \bibinfo {author}
  {\bibfnamefont {J.~C.}\ \bibnamefont {Stachowiak}}, \bibinfo {author}
  {\bibfnamefont {D.}~\bibnamefont {Stamou}}, \bibinfo {author} {\bibfnamefont
  {A.}~\bibnamefont {Breuer}}, \bibinfo {author} {\bibfnamefont
  {L.}~\bibnamefont {Lauritsen}}, \bibinfo {author} {\bibfnamefont
  {C.}~\bibnamefont {Simon}}, \bibinfo {author} {\bibfnamefont
  {C.}~\bibnamefont {Sykes}}, \bibinfo {author} {\bibfnamefont {G.~A.}\
  \bibnamefont {Voth}}, \ and\ \bibinfo {author} {\bibfnamefont {T.~R.}\
  \bibnamefont {Weikl}},\ }\href {https://doi.org/10.1088/1361-6463/aacb98}
  {\bibfield  {journal} {\bibinfo  {journal} {{J Phys D Appl Phys}}\ }\textbf
  {\bibinfo {volume} {51}} (\bibinfo {year} {2018})}\BibitemShut {NoStop}%
\bibitem [{\citenamefont {Traub}(2005)}]{Tra05}%
  \BibitemOpen
  \bibfield  {author} {\bibinfo {author} {\bibfnamefont {L.~M.}\ \bibnamefont
  {Traub}},\ }\href {\doibase 10.1016/j.bbamcr.2005.04.005} {\bibfield
  {journal} {\bibinfo  {journal} {Biochim Biophys Acta}\ }\textbf {\bibinfo
  {volume} {1744}},\ \bibinfo {pages} {415} (\bibinfo {year}
  {2005})}\BibitemShut {NoStop}%
\bibitem [{\citenamefont {Zamparo}\ \emph {et~al.}(2015)\citenamefont
  {Zamparo}, \citenamefont {Chianale}, \citenamefont {Tebaldi}, \citenamefont
  {Cosentino-Lagomarsino}, \citenamefont {Nicodemi},\ and\ \citenamefont
  {Gamba}}]{ZCT+15}%
  \BibitemOpen
  \bibfield  {author} {\bibinfo {author} {\bibfnamefont {M.}~\bibnamefont
  {Zamparo}}, \bibinfo {author} {\bibfnamefont {F.}~\bibnamefont {Chianale}},
  \bibinfo {author} {\bibfnamefont {C.}~\bibnamefont {Tebaldi}}, \bibinfo
  {author} {\bibfnamefont {M.}~\bibnamefont {Cosentino-Lagomarsino}}, \bibinfo
  {author} {\bibfnamefont {M.}~\bibnamefont {Nicodemi}}, \ and\ \bibinfo
  {author} {\bibfnamefont {A.}~\bibnamefont {Gamba}},\ }\href {\doibase
  10.1039/C4SM02157F} {\bibfield  {journal} {\bibinfo  {journal} {Soft Matter}\
  }\textbf {\bibinfo {volume} {11}},\ \bibinfo {pages} {838} (\bibinfo {year}
  {2015})}\BibitemShut {NoStop}%
\bibitem [{\citenamefont {Halatek}\ \emph {et~al.}(2018)\citenamefont
  {Halatek}, \citenamefont {Brauns},\ and\ \citenamefont {Frey}}]{HBF18}%
  \BibitemOpen
  \bibfield  {author} {\bibinfo {author} {\bibfnamefont {J.}~\bibnamefont
  {Halatek}}, \bibinfo {author} {\bibfnamefont {F.}~\bibnamefont {Brauns}}, \
  and\ \bibinfo {author} {\bibfnamefont {E.}~\bibnamefont {Frey}},\ }\href
  {\doibase 10.1098/rstb.2017.0107} {\bibfield  {journal} {\bibinfo  {journal}
  {Philosophical Transactions of the Royal Society B: Biological Sciences}\
  }\textbf {\bibinfo {volume} {373}},\ \bibinfo {pages} {20170107} (\bibinfo
  {year} {2018})}\BibitemShut {NoStop}%
\bibitem [{\citenamefont {Berry}\ \emph {et~al.}(2018)\citenamefont {Berry},
  \citenamefont {Brangwynne},\ and\ \citenamefont {Haataja}}]{BBH18}%
  \BibitemOpen
  \bibfield  {author} {\bibinfo {author} {\bibfnamefont {J.}~\bibnamefont
  {Berry}}, \bibinfo {author} {\bibfnamefont {C.~P.}\ \bibnamefont
  {Brangwynne}}, \ and\ \bibinfo {author} {\bibfnamefont {M.}~\bibnamefont
  {Haataja}},\ }\href {\doibase 10.1088/1361-6633/aaa61e} {\bibfield  {journal}
  {\bibinfo  {journal} {Reports on Progress in Physics}\ }\textbf {\bibinfo
  {volume} {81}},\ \bibinfo {pages} {046601} (\bibinfo {year}
  {2018})}\BibitemShut {NoStop}%
\bibitem [{\citenamefont {Lyon}\ \emph {et~al.}(2020)\citenamefont {Lyon},
  \citenamefont {Peeples},\ and\ \citenamefont {Rosen}}]{LPR20}%
  \BibitemOpen
  \bibfield  {author} {\bibinfo {author} {\bibfnamefont {A.~S.}\ \bibnamefont
  {Lyon}}, \bibinfo {author} {\bibfnamefont {W.~B.}\ \bibnamefont {Peeples}}, \
  and\ \bibinfo {author} {\bibfnamefont {M.~K.}\ \bibnamefont {Rosen}},\ }\href
  {https://doi.org/10.1038/s41580-020-00303-z} {\bibfield  {journal} {\bibinfo
  {journal} {{Nat Rev Mol Cell Biol}}\ } (\bibinfo {year} {2020})}\BibitemShut
  {NoStop}%
\bibitem [{\citenamefont {Liu}\ \emph {et~al.}(2010)\citenamefont {Liu},
  \citenamefont {Aguet}, \citenamefont {Danuser},\ and\ \citenamefont
  {Schmid}}]{LAD+10}%
  \BibitemOpen
  \bibfield  {author} {\bibinfo {author} {\bibfnamefont {A.~P.}\ \bibnamefont
  {Liu}}, \bibinfo {author} {\bibfnamefont {F.}~\bibnamefont {Aguet}}, \bibinfo
  {author} {\bibfnamefont {G.}~\bibnamefont {Danuser}}, \ and\ \bibinfo
  {author} {\bibfnamefont {S.~L.}\ \bibnamefont {Schmid}},\ }\href {\doibase
  10.1083/jcb.201008117} {\bibfield  {journal} {\bibinfo  {journal} {J Cell
  Biol}\ }\textbf {\bibinfo {volume} {191}},\ \bibinfo {pages} {1381} (\bibinfo
  {year} {2010})}\BibitemShut {NoStop}%
\bibitem [{\citenamefont {Puri}\ \emph {et~al.}(2005)\citenamefont {Puri},
  \citenamefont {Tosoni}, \citenamefont {Comai}, \citenamefont {Rabellino},
  \citenamefont {Segat}, \citenamefont {Caneva}, \citenamefont {Luzzi},
  \citenamefont {{Di Fiore}},\ and\ \citenamefont {Tacchetti}}]{PTC+05}%
  \BibitemOpen
  \bibfield  {author} {\bibinfo {author} {\bibfnamefont {C.}~\bibnamefont
  {Puri}}, \bibinfo {author} {\bibfnamefont {D.}~\bibnamefont {Tosoni}},
  \bibinfo {author} {\bibfnamefont {R.}~\bibnamefont {Comai}}, \bibinfo
  {author} {\bibfnamefont {A.}~\bibnamefont {Rabellino}}, \bibinfo {author}
  {\bibfnamefont {D.}~\bibnamefont {Segat}}, \bibinfo {author} {\bibfnamefont
  {F.}~\bibnamefont {Caneva}}, \bibinfo {author} {\bibfnamefont
  {P.}~\bibnamefont {Luzzi}}, \bibinfo {author} {\bibfnamefont {P.~P.}\
  \bibnamefont {{Di Fiore}}}, \ and\ \bibinfo {author} {\bibfnamefont
  {C.}~\bibnamefont {Tacchetti}},\ }\href {\doibase 10.1091/mbc.e04-07-0596}
  {\bibfield  {journal} {\bibinfo  {journal} {Molecular Biology of the Cell}\
  }\textbf {\bibinfo {volume} {16}},\ \bibinfo {pages} {2704} (\bibinfo {year}
  {2005})}\BibitemShut {NoStop}%
\bibitem [{\citenamefont {Sens}\ \emph {et~al.}(2008)\citenamefont {Sens},
  \citenamefont {Johannes},\ and\ \citenamefont {Bassereau}}]{SJB08}%
  \BibitemOpen
  \bibfield  {author} {\bibinfo {author} {\bibfnamefont {P.}~\bibnamefont
  {Sens}}, \bibinfo {author} {\bibfnamefont {L.}~\bibnamefont {Johannes}}, \
  and\ \bibinfo {author} {\bibfnamefont {P.}~\bibnamefont {Bassereau}},\ }\href
  {\doibase 10.1016/j.ceb.2008.04.004} {\bibfield  {journal} {\bibinfo
  {journal} {Curr. Opin. Cell Biol.}\ }\textbf {\bibinfo {volume} {20}},\
  \bibinfo {pages} {476} (\bibinfo {year} {2008})}\BibitemShut {NoStop}%
\bibitem [{\citenamefont {Busch}\ \emph {et~al.}(2015)\citenamefont {Busch},
  \citenamefont {Houser}, \citenamefont {Hayden}, \citenamefont {Sherman},
  \citenamefont {Lafer},\ and\ \citenamefont {Stachowiak}}]{BHH+15}%
  \BibitemOpen
  \bibfield  {author} {\bibinfo {author} {\bibfnamefont {D.~J.}\ \bibnamefont
  {Busch}}, \bibinfo {author} {\bibfnamefont {J.~R.}\ \bibnamefont {Houser}},
  \bibinfo {author} {\bibfnamefont {C.~C.}\ \bibnamefont {Hayden}}, \bibinfo
  {author} {\bibfnamefont {M.~B.}\ \bibnamefont {Sherman}}, \bibinfo {author}
  {\bibfnamefont {E.~M.}\ \bibnamefont {Lafer}}, \ and\ \bibinfo {author}
  {\bibfnamefont {J.~C.}\ \bibnamefont {Stachowiak}},\ }\href {\doibase
  10.1038/ncomms8875} {\bibfield  {journal} {\bibinfo  {journal} {Nat Commun}\
  }\textbf {\bibinfo {volume} {6}},\ \bibinfo {pages} {7875} (\bibinfo {year}
  {2015})}\BibitemShut {NoStop}%
\bibitem [{\citenamefont {Stachowiak}\ \emph {et~al.}(2012)\citenamefont
  {Stachowiak}, \citenamefont {Schmid}, \citenamefont {Ryan}, \citenamefont
  {Ann}, \citenamefont {Sasaki}, \citenamefont {Sherman}, \citenamefont
  {Geissler}, \citenamefont {Fletcher},\ and\ \citenamefont {Hayden}}]{SSR+12}%
  \BibitemOpen
  \bibfield  {author} {\bibinfo {author} {\bibfnamefont {J.~C.}\ \bibnamefont
  {Stachowiak}}, \bibinfo {author} {\bibfnamefont {E.~M.}\ \bibnamefont
  {Schmid}}, \bibinfo {author} {\bibfnamefont {C.~J.}\ \bibnamefont {Ryan}},
  \bibinfo {author} {\bibfnamefont {H.~S.}\ \bibnamefont {Ann}}, \bibinfo
  {author} {\bibfnamefont {D.~Y.}\ \bibnamefont {Sasaki}}, \bibinfo {author}
  {\bibfnamefont {M.~B.}\ \bibnamefont {Sherman}}, \bibinfo {author}
  {\bibfnamefont {P.~L.}\ \bibnamefont {Geissler}}, \bibinfo {author}
  {\bibfnamefont {D.~A.}\ \bibnamefont {Fletcher}}, \ and\ \bibinfo {author}
  {\bibfnamefont {C.~C.}\ \bibnamefont {Hayden}},\ }\href {\doibase
  10.1016/j.bbamem.2016.03.009} {\bibfield  {journal} {\bibinfo  {journal}
  {Nat. Cell Biol.}\ }\textbf {\bibinfo {volume} {14}},\ \bibinfo {pages} {944}
  (\bibinfo {year} {2012})}\BibitemShut {NoStop}%
\bibitem [{\citenamefont {Kaksonen}\ and\ \citenamefont {Roux}(2018)}]{KR18}%
  \BibitemOpen
  \bibfield  {author} {\bibinfo {author} {\bibfnamefont {M.}~\bibnamefont
  {Kaksonen}}\ and\ \bibinfo {author} {\bibfnamefont {A.}~\bibnamefont
  {Roux}},\ }\href {\doibase 10.1038/nrm.2017.132} {\bibfield  {journal}
  {\bibinfo  {journal} {Nat. Rev. Molec. Cell Biol.}\ }\textbf {\bibinfo
  {volume} {19}},\ \bibinfo {pages} {313} (\bibinfo {year} {2018})}\BibitemShut
  {NoStop}%
\bibitem [{\citenamefont {Gov}(2018)}]{Gov18}%
  \BibitemOpen
  \bibfield  {author} {\bibinfo {author} {\bibfnamefont {N.~S.}\ \bibnamefont
  {Gov}},\ }\href {\doibase 10.1098/rstb.2017.0115} {\bibfield  {journal}
  {\bibinfo  {journal} {Philosophical Transactions of the Royal Society B:
  Biological Sciences}\ }\textbf {\bibinfo {volume} {373}},\ \bibinfo {pages}
  {20170115} (\bibinfo {year} {2018})}\BibitemShut {NoStop}%
\bibitem [{\citenamefont {Chen}\ \emph {et~al.}(2016)\citenamefont {Chen},
  \citenamefont {Atefi},\ and\ \citenamefont {Baumgart}}]{CAB16}%
  \BibitemOpen
  \bibfield  {author} {\bibinfo {author} {\bibfnamefont {Z.}~\bibnamefont
  {Chen}}, \bibinfo {author} {\bibfnamefont {E.}~\bibnamefont {Atefi}}, \ and\
  \bibinfo {author} {\bibfnamefont {T.}~\bibnamefont {Baumgart}},\ }\href
  {\doibase 10.1016/j.bpj.2016.09.039} {\bibfield  {journal} {\bibinfo
  {journal} {{Biophys J}}\ }\textbf {\bibinfo {volume} {111}},\ \bibinfo
  {pages} {1823} (\bibinfo {year} {2016})}\BibitemShut {NoStop}%
\bibitem [{\citenamefont {Leibler}(1986)}]{Lei86}%
  \BibitemOpen
  \bibfield  {author} {\bibinfo {author} {\bibfnamefont {S.}~\bibnamefont
  {Leibler}},\ }\href {\doibase 10.1051/jphys:01986004703050700} {\bibfield
  {journal} {\bibinfo  {journal} {J. de Physique}\ }\textbf {\bibinfo {volume}
  {47}},\ \bibinfo {pages} {507} (\bibinfo {year} {1986})}\BibitemShut
  {NoStop}%
\bibitem [{\citenamefont {Leibler}\ and\ \citenamefont
  {Andelman}(1987)}]{LA87}%
  \BibitemOpen
  \bibfield  {author} {\bibinfo {author} {\bibfnamefont {S.}~\bibnamefont
  {Leibler}}\ and\ \bibinfo {author} {\bibfnamefont {D.}~\bibnamefont
  {Andelman}},\ }\href {\doibase 10.1051/jphys:0198700480110201300} {\bibfield
  {journal} {\bibinfo  {journal} {J. de Physique}\ }\textbf {\bibinfo {volume}
  {48}},\ \bibinfo {pages} {2013} (\bibinfo {year} {1987})}\BibitemShut
  {NoStop}%
\bibitem [{\citenamefont {Bitbol}\ \emph {et~al.}(2011)\citenamefont {Bitbol},
  \citenamefont {Peliti},\ and\ \citenamefont {Fournier}}]{BPF11}%
  \BibitemOpen
  \bibfield  {author} {\bibinfo {author} {\bibfnamefont {A.~F.}\ \bibnamefont
  {Bitbol}}, \bibinfo {author} {\bibfnamefont {L.}~\bibnamefont {Peliti}}, \
  and\ \bibinfo {author} {\bibfnamefont {J.~B.}\ \bibnamefont {Fournier}},\
  }\href {https://doi.org/10.1140/epje/i2011-11053-4} {\bibfield  {journal}
  {\bibinfo  {journal} {Europ. Phys. J. E}\ }\textbf {\bibinfo {volume} {34}}
  (\bibinfo {year} {2011})}\BibitemShut {NoStop}%
\bibitem [{\citenamefont {Banerjee}\ \emph {et~al.}(2012)\citenamefont
  {Banerjee}, \citenamefont {Berezhkovskii},\ and\ \citenamefont
  {Nossal}}]{BBN12}%
  \BibitemOpen
  \bibfield  {author} {\bibinfo {author} {\bibfnamefont {A.}~\bibnamefont
  {Banerjee}}, \bibinfo {author} {\bibfnamefont {A.}~\bibnamefont
  {Berezhkovskii}}, \ and\ \bibinfo {author} {\bibfnamefont {R.}~\bibnamefont
  {Nossal}},\ }\href {\doibase 10.1016/j.bpj.2012.05.010} {\bibfield  {journal}
  {\bibinfo  {journal} {Biophys. J.}\ }\textbf {\bibinfo {volume} {102}},\
  \bibinfo {pages} {2725} (\bibinfo {year} {2012})}\BibitemShut {NoStop}%
\bibitem [{\citenamefont {Rautu}\ \emph {et~al.}(2015)\citenamefont {Rautu},
  \citenamefont {Rowlands},\ and\ \citenamefont {Turner}}]{RRT15}%
  \BibitemOpen
  \bibfield  {author} {\bibinfo {author} {\bibfnamefont {S.~A.}\ \bibnamefont
  {Rautu}}, \bibinfo {author} {\bibfnamefont {G.}~\bibnamefont {Rowlands}}, \
  and\ \bibinfo {author} {\bibfnamefont {M.~S.}\ \bibnamefont {Turner}},\
  }\href {\doibase 10.1103/physrevlett.114.098101} {\bibfield  {journal}
  {\bibinfo  {journal} {Physical Review Letters}\ }\textbf {\bibinfo {volume}
  {114}},\ \bibinfo {pages} {098101} (\bibinfo {year} {2015})}\BibitemShut
  {NoStop}%
\bibitem [{\citenamefont {Foret}\ and\ \citenamefont {Sens}(2008)}]{FS08a}%
  \BibitemOpen
  \bibfield  {author} {\bibinfo {author} {\bibfnamefont {L.}~\bibnamefont
  {Foret}}\ and\ \bibinfo {author} {\bibfnamefont {P.}~\bibnamefont {Sens}},\
  }\href {https://doi-org.proxy.unimib.it/10.1073/pnas.0801173105} {\bibfield
  {journal} {\bibinfo  {journal} {Proc. Natl. Acad. Sci. USA}\ }\textbf
  {\bibinfo {volume} {105}},\ \bibinfo {pages} {14763} (\bibinfo {year}
  {2008})}\BibitemShut {NoStop}%
\bibitem [{\citenamefont {Barabasi}\ and\ \citenamefont
  {Stanley}(1995)}]{BS95}%
  \BibitemOpen
  \bibfield  {author} {\bibinfo {author} {\bibfnamefont {A.-L.}\ \bibnamefont
  {Barabasi}}\ and\ \bibinfo {author} {\bibfnamefont {H.~E.}\ \bibnamefont
  {Stanley}},\ }\href@noop {} {\emph {\bibinfo {title} {Fractal Concepts in
  Surface Growth}}}\ (\bibinfo  {publisher} {Cambridge University Press,
  Cambridge},\ \bibinfo {year} {1995})\BibitemShut {NoStop}%
\bibitem [{\citenamefont {{Hele-Shaw}}(1898)}]{HS98}%
  \BibitemOpen
  \bibfield  {author} {\bibinfo {author} {\bibfnamefont {H.~S.}\ \bibnamefont
  {{Hele-Shaw}}},\ }\href {\doibase 10.1038/058034a0} {\bibfield  {journal}
  {\bibinfo  {journal} {Nature}\ }\textbf {\bibinfo {volume} {58}},\ \bibinfo
  {pages} {34} (\bibinfo {year} {1898})}\BibitemShut {NoStop}%
\bibitem [{\citenamefont {Slezov}(2009)}]{Sle09}%
  \BibitemOpen
  \bibfield  {author} {\bibinfo {author} {\bibfnamefont {V.~V.}\ \bibnamefont
  {Slezov}},\ }\href@noop {} {\emph {\bibinfo {title} {Kinetics of First Order
  Phase Transitions}}}\ (\bibinfo  {publisher} {Wiley-VCH Verlag GmbH \& Co.
  KGaA, Weinheim},\ \bibinfo {year} {2009})\BibitemShut {NoStop}%
\bibitem [{\citenamefont {Gamba}\ \emph {et~al.}(2007)\citenamefont {Gamba},
  \citenamefont {Kolokolov}, \citenamefont {Lebedev},\ and\ \citenamefont
  {Ortenzi}}]{GKL+07}%
  \BibitemOpen
  \bibfield  {author} {\bibinfo {author} {\bibfnamefont {A.}~\bibnamefont
  {Gamba}}, \bibinfo {author} {\bibfnamefont {I.}~\bibnamefont {Kolokolov}},
  \bibinfo {author} {\bibfnamefont {V.}~\bibnamefont {Lebedev}}, \ and\
  \bibinfo {author} {\bibfnamefont {G.}~\bibnamefont {Ortenzi}},\ }\href
  {\doibase 10.1103/PhysRevLett.99.158101} {\bibfield  {journal} {\bibinfo
  {journal} {Phys Rev Lett}\ }\textbf {\bibinfo {volume} {99}},\ \bibinfo
  {pages} {158101} (\bibinfo {year} {2007})}\BibitemShut {NoStop}%
\bibitem [{\citenamefont {Bertini}\ \emph {et~al.}(2007)\citenamefont
  {Bertini}, \citenamefont {Sole}, \citenamefont {Gabrielli}, \citenamefont
  {Jona-Lasinio},\ and\ \citenamefont {Landim}}]{BSG+07}%
  \BibitemOpen
  \bibfield  {author} {\bibinfo {author} {\bibfnamefont {L.}~\bibnamefont
  {Bertini}}, \bibinfo {author} {\bibfnamefont {A.~D.}\ \bibnamefont {Sole}},
  \bibinfo {author} {\bibfnamefont {D.}~\bibnamefont {Gabrielli}}, \bibinfo
  {author} {\bibfnamefont {G.}~\bibnamefont {Jona-Lasinio}}, \ and\ \bibinfo
  {author} {\bibfnamefont {C.}~\bibnamefont {Landim}},\ }\href {\doibase
  10.1088/1742-5468/2007/07/p07014} {\bibfield  {journal} {\bibinfo  {journal}
  {J. Stat. Mech.: Theory and Experiment}\ }\textbf {\bibinfo {volume}
  {2007}},\ \bibinfo {pages} {P07014} (\bibinfo {year} {2007})}\BibitemShut
  {NoStop}%
\bibitem [{\citenamefont {Zamparo}\ \emph {et~al.}(2019)\citenamefont
  {Zamparo}, \citenamefont {Dall'Asta},\ and\ \citenamefont {Gamba}}]{ZAG19}%
  \BibitemOpen
  \bibfield  {author} {\bibinfo {author} {\bibfnamefont {M.}~\bibnamefont
  {Zamparo}}, \bibinfo {author} {\bibfnamefont {L.}~\bibnamefont {Dall'Asta}},
  \ and\ \bibinfo {author} {\bibfnamefont {A.}~\bibnamefont {Gamba}},\ }\href
  {https:/doi.org/10.1007/s10955-018-2175-x} {\bibfield  {journal} {\bibinfo
  {journal} {J. Stat. Phys.}\ }\textbf {\bibinfo {volume} {174}},\ \bibinfo
  {pages} {120} (\bibinfo {year} {2019})}\BibitemShut {NoStop}%
\bibitem [{Note1()}]{Note1}%
  \BibitemOpen
  \bibinfo {note} {Since $\phi $ is a monotonically increasing function of the
  insertion rate~$k_I$~(see SM), by a change of variable~$\phi $ (which is
  directly observable) can be used as a control parameter in the place of~$k_I$
  (which is not observable).}\BibitemShut {Stop}%
\bibitem [{\citenamefont {Goldstein}\ \emph {et~al.}(1981)\citenamefont
  {Goldstein}, \citenamefont {Wofsy},\ and\ \citenamefont {Bell}}]{GWB81}%
  \BibitemOpen
  \bibfield  {author} {\bibinfo {author} {\bibfnamefont {B.}~\bibnamefont
  {Goldstein}}, \bibinfo {author} {\bibfnamefont {C.}~\bibnamefont {Wofsy}}, \
  and\ \bibinfo {author} {\bibfnamefont {G.}~\bibnamefont {Bell}},\ }\href
  {\doibase 10.1073/pnas.78.9.5695} {\bibfield  {journal} {\bibinfo  {journal}
  {Proc. Natl. Acad. Sci. USA}\ }\textbf {\bibinfo {volume} {78}},\ \bibinfo
  {pages} {5695} (\bibinfo {year} {1981})}\BibitemShut {NoStop}%
\bibitem [{\citenamefont {Gillespie}(1976)}]{Gil76}%
  \BibitemOpen
  \bibfield  {author} {\bibinfo {author} {\bibfnamefont {D.~T.}\ \bibnamefont
  {Gillespie}},\ }\href {\doibase 10.1016/0021-9991(76)90041-3} {\bibfield
  {journal} {\bibinfo  {journal} {J. Comput. Phys.}\ }\textbf {\bibinfo
  {volume} {22}},\ \bibinfo {pages} {403} (\bibinfo {year} {1976})}\BibitemShut
  {NoStop}%
\bibitem [{\citenamefont {Brown}\ \emph {et~al.}(1986)\citenamefont {Brown},
  \citenamefont {Goldstein} \emph {et~al.}}]{BG86}%
  \BibitemOpen
  \bibfield  {author} {\bibinfo {author} {\bibfnamefont {M.~S.}\ \bibnamefont
  {Brown}}, \bibinfo {author} {\bibfnamefont {J.~L.}\ \bibnamefont
  {Goldstein}},  \emph {et~al.},\ }\href
  {https://doi.org/10.1126/science.3513311} {\bibfield  {journal} {\bibinfo
  {journal} {Science}\ }\textbf {\bibinfo {volume} {232}},\ \bibinfo {pages}
  {34} (\bibinfo {year} {1986})}\BibitemShut {NoStop}%
\bibitem [{\citenamefont {Hevonoja}\ \emph {et~al.}(2000)\citenamefont
  {Hevonoja}, \citenamefont {Pentik{\"a}inen}, \citenamefont {Hyv{\"o}nen},
  \citenamefont {Kovanen},\ and\ \citenamefont {Ala-Korpela}}]{HPH+00}%
  \BibitemOpen
  \bibfield  {author} {\bibinfo {author} {\bibfnamefont {T.}~\bibnamefont
  {Hevonoja}}, \bibinfo {author} {\bibfnamefont {M.~O.}\ \bibnamefont
  {Pentik{\"a}inen}}, \bibinfo {author} {\bibfnamefont {M.~T.}\ \bibnamefont
  {Hyv{\"o}nen}}, \bibinfo {author} {\bibfnamefont {P.~T.}\ \bibnamefont
  {Kovanen}}, \ and\ \bibinfo {author} {\bibfnamefont {M.}~\bibnamefont
  {Ala-Korpela}},\ }\href {\doibase 10.1016/S1388-1981(00)00123-2} {\bibfield
  {journal} {\bibinfo  {journal} {Biochimica et Biophysica Acta (BBA) -
  Molecular and Cell Biology of Lipids}\ }\textbf {\bibinfo {volume} {1488}},\
  \bibinfo {pages} {189} (\bibinfo {year} {2000})}\BibitemShut {NoStop}%
\bibitem [{\citenamefont {Schneider}(2016)}]{Sch16}%
  \BibitemOpen
  \bibfield  {author} {\bibinfo {author} {\bibfnamefont {W.~J.}\ \bibnamefont
  {Schneider}},\ }in\ \href {\doibase 10.1016/B978-0-444-63438-2.00017-1}
  {\emph {\bibinfo {booktitle} {Biochemistry of Lipids, Lipoproteins and
  Membranes (Sixth Edition)}}},\ \bibinfo {editor} {edited by\ \bibinfo
  {editor} {\bibfnamefont {N.~D.}\ \bibnamefont {Ridgway}}\ and\ \bibinfo
  {editor} {\bibfnamefont {R.~S.}\ \bibnamefont {McLeod}}}\ (\bibinfo
  {publisher} {Elsevier},\ \bibinfo {year} {2016})\ \bibinfo {edition} {6th
  }\ ed \BibitemShut {NoStop}%
\bibitem [{\citenamefont {Innerarity}\ \emph {et~al.}(1986)\citenamefont
  {Innerarity}, \citenamefont {Pitas},\ and\ \citenamefont {Mahley}}]{IPM86}%
  \BibitemOpen
  \bibfield  {author} {\bibinfo {author} {\bibfnamefont {T.~L.}\ \bibnamefont
  {Innerarity}}, \bibinfo {author} {\bibfnamefont {R.~E.}\ \bibnamefont
  {Pitas}}, \ and\ \bibinfo {author} {\bibfnamefont {R.~W.}\ \bibnamefont
  {Mahley}},\ }in\ \href {\doibase 10.1016/0076-6879(86)29091-6} {\emph
  {\bibinfo {booktitle} {Plasma Lipoproteins Part B: Characterization, Cell
  Biology, and Metabolism}}},\ \bibinfo {series} {Methods in Enzymology}, Vol.\
  \bibinfo {volume} {129}\ (\bibinfo  {publisher} {Academic Press},\ \bibinfo
  {year} {1986})\BibitemShut {NoStop}%
\bibitem [{\citenamefont {Ehrlich}\ \emph {et~al.}(2004)\citenamefont
  {Ehrlich}, \citenamefont {Boll}, \citenamefont {{Van Oijen}}, \citenamefont
  {Hariharan}, \citenamefont {Chandran}, \citenamefont {Nibert},\ and\
  \citenamefont {Kirchhausen}}]{EBV+04}%
  \BibitemOpen
  \bibfield  {author} {\bibinfo {author} {\bibfnamefont {M.}~\bibnamefont
  {Ehrlich}}, \bibinfo {author} {\bibfnamefont {W.}~\bibnamefont {Boll}},
  \bibinfo {author} {\bibfnamefont {A.}~\bibnamefont {{Van Oijen}}}, \bibinfo
  {author} {\bibfnamefont {R.}~\bibnamefont {Hariharan}}, \bibinfo {author}
  {\bibfnamefont {K.}~\bibnamefont {Chandran}}, \bibinfo {author}
  {\bibfnamefont {M.~L.}\ \bibnamefont {Nibert}}, \ and\ \bibinfo {author}
  {\bibfnamefont {T.}~\bibnamefont {Kirchhausen}},\ }\href {\doibase
  10.1016/j.cell.2004.08.017} {\bibfield  {journal} {\bibinfo  {journal}
  {Cell}\ }\textbf {\bibinfo {volume} {118}},\ \bibinfo {pages} {591} (\bibinfo
  {year} {2004})}\BibitemShut {NoStop}%
\bibitem [{\citenamefont {Oheim}\ \emph {et~al.}(2019)\citenamefont {Oheim},
  \citenamefont {Salomon}, \citenamefont {Weissman}, \citenamefont
  {Brunstein},\ and\ \citenamefont {Becherer}}]{OSW+19}%
  \BibitemOpen
  \bibfield  {author} {\bibinfo {author} {\bibfnamefont {M.}~\bibnamefont
  {Oheim}}, \bibinfo {author} {\bibfnamefont {A.}~\bibnamefont {Salomon}},
  \bibinfo {author} {\bibfnamefont {A.}~\bibnamefont {Weissman}}, \bibinfo
  {author} {\bibfnamefont {M.}~\bibnamefont {Brunstein}}, \ and\ \bibinfo
  {author} {\bibfnamefont {U.}~\bibnamefont {Becherer}},\ }\href {\doibase
  10.1016/j.bpj.2019.07.048} {\bibfield  {journal} {\bibinfo  {journal}
  {Biophysical Journal}\ }\textbf {\bibinfo {volume} {117}},\ \bibinfo {pages}
  {795} (\bibinfo {year} {2019})}\BibitemShut {NoStop}%
\bibitem [{\citenamefont {Waters}\ and\ \citenamefont {(Eds.)}(2014)}]{WW14}%
  \BibitemOpen
  \bibfield  {author} {\bibinfo {author} {\bibfnamefont {J.~C.}\ \bibnamefont
  {Waters}}\ and\ \bibinfo {author} {\bibfnamefont {T.~W.}\ \bibnamefont
  {(Eds.)}},\ }\href@noop {} {\emph {\bibinfo {title} {Quantitative Imaging in
  Cell Biology}}},\ \bibinfo {edition} {1st}\ ed.,\ Methods in Cell Biology
  Volume 123\ (\bibinfo  {publisher} {Academic Press},\ \bibinfo {year}
  {2014})\BibitemShut {NoStop}%
\bibitem [{\citenamefont {Khaw}\ \emph {et~al.}(2018)\citenamefont {Khaw},
  \citenamefont {Croop}, \citenamefont {Tang}, \citenamefont {M{\"o}hl},
  \citenamefont {Fuchs},\ and\ \citenamefont {Han}}]{KCT+18}%
  \BibitemOpen
  \bibfield  {author} {\bibinfo {author} {\bibfnamefont {I.}~\bibnamefont
  {Khaw}}, \bibinfo {author} {\bibfnamefont {B.}~\bibnamefont {Croop}},
  \bibinfo {author} {\bibfnamefont {J.}~\bibnamefont {Tang}}, \bibinfo {author}
  {\bibfnamefont {A.}~\bibnamefont {M{\"o}hl}}, \bibinfo {author} {\bibfnamefont
  {U.}~\bibnamefont {Fuchs}}, \ and\ \bibinfo {author} {\bibfnamefont {K.~Y.}\
  \bibnamefont {Han}},\ }\href {\doibase 10.1364/oe.26.015276} {\bibfield
  {journal} {\bibinfo  {journal} {Optics Express}\ }\textbf {\bibinfo {volume}
  {26}},\ \bibinfo {pages} {15276} (\bibinfo {year} {2018})}\BibitemShut
  {NoStop}%
\bibitem [{\citenamefont {Model}(2006)}]{Mod06}%
  \BibitemOpen
  \bibfield  {author} {\bibinfo {author} {\bibfnamefont {M.~A.}\ \bibnamefont
  {Model}},\ }\href {\doibase 10.1002/0471142956.cy1014s37} {\bibfield
  {journal} {\bibinfo  {journal} {Current Protocols in Cytometry}\ }\textbf
  {\bibinfo {volume} {37}},\ \bibinfo {pages} {10.14.1} (\bibinfo {year}
  {2006})}\BibitemShut {NoStop}%
\bibitem [{\citenamefont {Lindeberg}(1998)}]{Lin98}%
  \BibitemOpen
  \bibfield  {author} {\bibinfo {author} {\bibfnamefont {T.}~\bibnamefont
  {Lindeberg}},\ }\href {https://doi.org/10.1023/A:1008045108935} {\bibfield
  {journal} {\bibinfo  {journal} {Int. J. of computer vision}\ }\textbf
  {\bibinfo {volume} {30}},\ \bibinfo {pages} {79} (\bibinfo {year}
  {1998})}\BibitemShut {NoStop}%
\bibitem [{\citenamefont {Dekking}\ \emph {et~al.}(2007)\citenamefont
  {Dekking}, \citenamefont {Kraaikamp}, \citenamefont {Lopuha{\"a}},\ and\
  \citenamefont {Meester}}]{DKL+07}%
  \BibitemOpen
  \bibfield  {author} {\bibinfo {author} {\bibfnamefont {F.}~\bibnamefont
  {Dekking}}, \bibinfo {author} {\bibfnamefont {C.}~\bibnamefont {Kraaikamp}},
  \bibinfo {author} {\bibfnamefont {H.}~\bibnamefont {Lopuha{\"a}}}, \ and\
  \bibinfo {author} {\bibfnamefont {L.}~\bibnamefont {Meester}},\ }\href@noop
  {} {\emph {\bibinfo {title} {A Modern Introduction to Probability and
  Statistics: Understanding why and how}}}\ (\bibinfo  {publisher} {Springer},\
  \bibinfo {year} {2007})\BibitemShut {NoStop}%
\bibitem [{\citenamefont {Dempsey}\ \emph {et~al.}(2011)\citenamefont
  {Dempsey}, \citenamefont {Vaughan}, \citenamefont {Chen}, \citenamefont
  {Bates},\ and\ \citenamefont {Zhuang}}]{DVC+11}%
  \BibitemOpen
  \bibfield  {author} {\bibinfo {author} {\bibfnamefont {G.~T.}\ \bibnamefont
  {Dempsey}}, \bibinfo {author} {\bibfnamefont {J.~C.}\ \bibnamefont
  {Vaughan}}, \bibinfo {author} {\bibfnamefont {K.~H.}\ \bibnamefont {Chen}},
  \bibinfo {author} {\bibfnamefont {M.}~\bibnamefont {Bates}}, \ and\ \bibinfo
  {author} {\bibfnamefont {X.}~\bibnamefont {Zhuang}},\ }\href
  {https://doi.org/10.1038/nmeth.1768} {\bibfield  {journal} {\bibinfo
  {journal} {Nature methods}\ }\textbf {\bibinfo {volume} {8}},\ \bibinfo
  {pages} {1027} (\bibinfo {year} {2011})}\BibitemShut {NoStop}%
\bibitem [{\citenamefont {Avinoam}\ \emph {et~al.}(2015)\citenamefont
  {Avinoam}, \citenamefont {Schorb}, \citenamefont {Beese}, \citenamefont
  {Briggs},\ and\ \citenamefont {Kaksonen}}]{ASB+15}%
  \BibitemOpen
  \bibfield  {author} {\bibinfo {author} {\bibfnamefont {O.}~\bibnamefont
  {Avinoam}}, \bibinfo {author} {\bibfnamefont {M.}~\bibnamefont {Schorb}},
  \bibinfo {author} {\bibfnamefont {C.~J.}\ \bibnamefont {Beese}}, \bibinfo
  {author} {\bibfnamefont {J.~A.~G.}\ \bibnamefont {Briggs}}, \ and\ \bibinfo
  {author} {\bibfnamefont {M.}~\bibnamefont {Kaksonen}},\ }\href {\doibase
  10.1126/science.aaa9555} {\bibfield  {journal} {\bibinfo  {journal}
  {Science}\ }\textbf {\bibinfo {volume} {348}},\ \bibinfo {pages} {1369}
  (\bibinfo {year} {2015})}\BibitemShut {NoStop}%
\bibitem [{\citenamefont {Kolb}\ \emph {et~al.}(1983)\citenamefont {Kolb},
  \citenamefont {Botet},\ and\ \citenamefont {Jullien}}]{KBJ83}%
  \BibitemOpen
  \bibfield  {author} {\bibinfo {author} {\bibfnamefont {M.}~\bibnamefont
  {Kolb}}, \bibinfo {author} {\bibfnamefont {R.}~\bibnamefont {Botet}}, \ and\
  \bibinfo {author} {\bibfnamefont {R.}~\bibnamefont {Jullien}},\ }\href
  {https://doi.org/10.1103/PhysRevLett.51.1123} {\bibfield  {journal} {\bibinfo
   {journal} {Phys. Rev. Lett.}\ }\textbf {\bibinfo {volume} {51}},\ \bibinfo
  {pages} {1123} (\bibinfo {year} {1983})}\BibitemShut {NoStop}%
\bibitem [{\citenamefont {Witten}\ and\ \citenamefont {Sander}(1981)}]{WS81}%
  \BibitemOpen
  \bibfield  {author} {\bibinfo {author} {\bibfnamefont {T.~A.}\ \bibnamefont
  {Witten}}\ and\ \bibinfo {author} {\bibfnamefont {L.~M.}\ \bibnamefont
  {Sander}},\ }\href {\doibase 10.1103/PhysRevLett.47.1400} {\bibfield
  {journal} {\bibinfo  {journal} {Phys. Rev. Lett.}\ }\textbf {\bibinfo
  {volume} {47}},\ \bibinfo {pages} {1400} (\bibinfo {year}
  {1981})}\BibitemShut {NoStop}%
\bibitem [{\citenamefont {Meakin}(1983)}]{Mea83}%
  \BibitemOpen
  \bibfield  {author} {\bibinfo {author} {\bibfnamefont {P.}~\bibnamefont
  {Meakin}},\ }\href {https://doi.org/10.1103/PhysRevLett.51.1119} {\bibfield
  {journal} {\bibinfo  {journal} {Phys. Rev. Lett.}\ }\textbf {\bibinfo
  {volume} {51}},\ \bibinfo {pages} {1119} (\bibinfo {year}
  {1983})}\BibitemShut {NoStop}%
\bibitem [{\citenamefont {Mana}\ \emph {et~al.}(2016)\citenamefont {Mana},
  \citenamefont {Clapero}, \citenamefont {Panieri}, \citenamefont {Panero},
  \citenamefont {B{\"o}ttcher}, \citenamefont {Tseng}, \citenamefont {Saltarin},
  \citenamefont {Astanina}, \citenamefont {Wolanska}, \citenamefont {Morgan},
  \citenamefont {Humphries}, \citenamefont {Santoro}, \citenamefont {Serini},\
  and\ \citenamefont {Valdembri}}]{MCP+16}%
  \BibitemOpen
  \bibfield  {author} {\bibinfo {author} {\bibfnamefont {G.}~\bibnamefont
  {Mana}}, \bibinfo {author} {\bibfnamefont {F.}~\bibnamefont {Clapero}},
  \bibinfo {author} {\bibfnamefont {E.}~\bibnamefont {Panieri}}, \bibinfo
  {author} {\bibfnamefont {V.}~\bibnamefont {Panero}}, \bibinfo {author}
  {\bibfnamefont {R.~T.}\ \bibnamefont {B{\"o}ttcher}}, \bibinfo {author}
  {\bibfnamefont {H.-Y.}\ \bibnamefont {Tseng}}, \bibinfo {author}
  {\bibfnamefont {F.}~\bibnamefont {Saltarin}}, \bibinfo {author}
  {\bibfnamefont {E.}~\bibnamefont {Astanina}}, \bibinfo {author}
  {\bibfnamefont {K.~I.}\ \bibnamefont {Wolanska}}, \bibinfo {author}
  {\bibfnamefont {M.~R.}\ \bibnamefont {Morgan}}, \bibinfo {author}
  {\bibfnamefont {M.}~\bibnamefont {Humphries}}, \bibinfo {author}
  {\bibfnamefont {M.}~\bibnamefont {Santoro}}, \bibinfo {author} {\bibfnamefont
  {G.}~\bibnamefont {Serini}}, \ and\ \bibinfo {author} {\bibfnamefont
  {D.}~\bibnamefont {Valdembri}},\ }\href {\doibase 10.1038/ncomms13546}
  {\bibfield  {journal} {\bibinfo  {journal} {Nat. Commun.}\ }\textbf {\bibinfo
  {volume} {7}},\ \bibinfo {pages} {13546} (\bibinfo {year}
  {2016})}\BibitemShut {NoStop}%
\bibitem [{\citenamefont {Picco}\ and\ \citenamefont {Kaksonen}(2018)}]{PK18}%
  \BibitemOpen
  \bibfield  {author} {\bibinfo {author} {\bibfnamefont {A.}~\bibnamefont
  {Picco}}\ and\ \bibinfo {author} {\bibfnamefont {M.}~\bibnamefont
  {Kaksonen}},\ }\href {\doibase 10.1016/j.ceb.2018.06.005} {\bibfield
  {journal} {\bibinfo  {journal} {Curr. Opin. Cell Biol.}\ }\textbf {\bibinfo
  {volume} {53}},\ \bibinfo {pages} {105} (\bibinfo {year} {2018})}\BibitemShut
  {NoStop}%
\bibitem [{\citenamefont {Barak}\ and\ \citenamefont {Webb}(1982)}]{BW82}%
  \BibitemOpen
  \bibfield  {author} {\bibinfo {author} {\bibfnamefont {L.~S.}\ \bibnamefont
  {Barak}}\ and\ \bibinfo {author} {\bibfnamefont {W.~W.}\ \bibnamefont
  {Webb}},\ }\href {\doibase 10.1083/jcb.95.3.846} {\bibfield  {journal}
  {\bibinfo  {journal} {J. Cell Biol.}\ }\textbf {\bibinfo {volume} {95}},\
  \bibinfo {pages} {846} (\bibinfo {year} {1982})}\BibitemShut {NoStop}%
\bibitem [{Note2()}]{Note2}%
  \BibitemOpen
  \bibinfo {note} {Single molecule measurements show that endocytic vesicles
  host $\pi R_E^2/A_0\sim 10$ LDL molecules~\cite {ABG77}.}\BibitemShut {Stop}%
\bibitem [{\citenamefont {S{\"o}nnichsen}\ \emph {et~al.}(2000)\citenamefont
  {Sönnichsen}, \citenamefont {{De Renzis}}, \citenamefont {Nielsen},
  \citenamefont {Rietdorf},\ and\ \citenamefont {Zerial}}]{SDN+00}%
  \BibitemOpen
  \bibfield  {author} {\bibinfo {author} {\bibfnamefont {B.}~\bibnamefont
  {S{\"o}nnichsen}}, \bibinfo {author} {\bibfnamefont {S.}~\bibnamefont {{De
  Renzis}}}, \bibinfo {author} {\bibfnamefont {E.}~\bibnamefont {Nielsen}},
  \bibinfo {author} {\bibfnamefont {J.}~\bibnamefont {Rietdorf}}, \ and\
  \bibinfo {author} {\bibfnamefont {M.}~\bibnamefont {Zerial}},\ }\href
  {\doibase 10.1083/jcb.149.4.901} {\bibfield  {journal} {\bibinfo  {journal}
  {J. Cell Biol.}\ }\textbf {\bibinfo {volume} {149}},\ \bibinfo {pages} {901}
  (\bibinfo {year} {2000})}\BibitemShut {NoStop}%
\bibitem [{\citenamefont {Wedlich-S{\"o}ldner}\ and\ \citenamefont
  {Li}(2003)}]{WL03}%
  \BibitemOpen
  \bibfield  {author} {\bibinfo {author} {\bibfnamefont {R.}~\bibnamefont
  {Wedlich-S{\"o}ldner}}\ and\ \bibinfo {author} {\bibfnamefont {R.}~\bibnamefont
  {Li}},\ }\href {\doibase 10.1038/ncb0403-267} {\bibfield  {journal} {\bibinfo
   {journal} {Nat. Cell Biol.}\ ,\ \bibinfo {pages} {267}} (\bibinfo {year}
  {2003})}\BibitemShut {NoStop}%
\bibitem [{\citenamefont {Rodriguez-Boulan}\ and\ \citenamefont
  {Macara}(2014)}]{RM14a}%
  \BibitemOpen
  \bibfield  {author} {\bibinfo {author} {\bibfnamefont {E.}~\bibnamefont
  {Rodriguez-Boulan}}\ and\ \bibinfo {author} {\bibfnamefont {I.~G.}\
  \bibnamefont {Macara}},\ }\href {https://doi.org/10.1038/nrm3775} {\bibfield
  {journal} {\bibinfo  {journal} {Nat. Rev. Molec. Cell Biol.}\ }\textbf
  {\bibinfo {volume} {15}},\ \bibinfo {pages} {225} (\bibinfo {year}
  {2014})}\BibitemShut {NoStop}%
\bibitem [{\citenamefont {Shewan}\ \emph {et~al.}(2011)\citenamefont {Shewan},
  \citenamefont {Eastburn},\ and\ \citenamefont {Mostov}}]{SEM11}%
  \BibitemOpen
  \bibfield  {author} {\bibinfo {author} {\bibfnamefont {A.}~\bibnamefont
  {Shewan}}, \bibinfo {author} {\bibfnamefont {D.~J.}\ \bibnamefont
  {Eastburn}}, \ and\ \bibinfo {author} {\bibfnamefont {K.}~\bibnamefont
  {Mostov}},\ }\href {\doibase 10.1101/cshperspect.a004796} {\bibfield
  {journal} {\bibinfo  {journal} {Cold Spring Harb. Perspect. Biol.}\ ,\
  \bibinfo {pages} {a004796}} (\bibinfo {year} {2011})}\BibitemShut {NoStop}%
\bibitem [{\citenamefont {Gamba}\ \emph {et~al.}(2005)\citenamefont {Gamba},
  \citenamefont {de~Candia}, \citenamefont {{Di Talia}}, \citenamefont
  {Coniglio}, \citenamefont {Bussolino},\ and\ \citenamefont
  {Serini}}]{GCT+05}%
  \BibitemOpen
  \bibfield  {author} {\bibinfo {author} {\bibfnamefont {A.}~\bibnamefont
  {Gamba}}, \bibinfo {author} {\bibfnamefont {A.}~\bibnamefont {de~Candia}},
  \bibinfo {author} {\bibfnamefont {S.}~\bibnamefont {{Di Talia}}}, \bibinfo
  {author} {\bibfnamefont {A.}~\bibnamefont {Coniglio}}, \bibinfo {author}
  {\bibfnamefont {F.}~\bibnamefont {Bussolino}}, \ and\ \bibinfo {author}
  {\bibfnamefont {G.}~\bibnamefont {Serini}},\ }\href {\doibase
  10.1073/pnas.0503974102} {\bibfield  {journal} {\bibinfo  {journal} {Proc.
  Natl. Acad. Sci. USA}\ }\textbf {\bibinfo {volume} {102}},\ \bibinfo {pages}
  {16927} (\bibinfo {year} {2005})}\BibitemShut {NoStop}%
\bibitem [{\citenamefont {Heinzer}\ \emph {et~al.}(2008)\citenamefont
  {Heinzer}, \citenamefont {Worz}, \citenamefont {Kalla}, \citenamefont
  {Rohr},\ and\ \citenamefont {Weiss}}]{HWK+08}%
  \BibitemOpen
  \bibfield  {author} {\bibinfo {author} {\bibfnamefont {S.}~\bibnamefont
  {Heinzer}}, \bibinfo {author} {\bibfnamefont {S.}~\bibnamefont {Worz}},
  \bibinfo {author} {\bibfnamefont {C.}~\bibnamefont {Kalla}}, \bibinfo
  {author} {\bibfnamefont {K.}~\bibnamefont {Rohr}}, \ and\ \bibinfo {author}
  {\bibfnamefont {M.}~\bibnamefont {Weiss}},\ }\href {\doibase
  10.1242/jcs.013383} {\bibfield  {journal} {\bibinfo  {journal} {J. Cell
  Sci.}\ }\textbf {\bibinfo {volume} {121}},\ \bibinfo {pages} {55} (\bibinfo
  {year} {2008})}\BibitemShut {NoStop}%
\bibitem [{\citenamefont {Anderson}\ \emph {et~al.}(1977)\citenamefont
  {Anderson}, \citenamefont {Brown},\ and\ \citenamefont {Goldstein}}]{ABG77}%
  \BibitemOpen
  \bibfield  {author} {\bibinfo {author} {\bibfnamefont {R.~G.}\ \bibnamefont
  {Anderson}}, \bibinfo {author} {\bibfnamefont {M.~S.}\ \bibnamefont {Brown}},
  \ and\ \bibinfo {author} {\bibfnamefont {J.~L.}\ \bibnamefont {Goldstein}},\
  }\href {https://doi.org/10.1016/0092-8674(77)90022-8} {\bibfield  {journal}
  {\bibinfo  {journal} {Cell}\ }\textbf {\bibinfo {volume} {10}},\ \bibinfo
  {pages} {351} (\bibinfo {year} {1977})}\BibitemShut {NoStop}%
\end{thebibliography}


%
\end{document}